\RequirePackage{silence}
\documentclass{optica-article}
\journal{opticajournal} 
\articletype{Research Article}
\usepackage[justification=justified,singlelinecheck=true]{caption}
\usepackage{lineno}
\usepackage{float} 
\usepackage{siunitx}
\usepackage{caption}
\usepackage{float}
\usepackage{placeins}
\usepackage{amsmath}
\begin{document}



\title{Gaussian process regression-based inverse estimation of strain and temperature using phase shift fiber cavity ring down spectroscopy}
\author{M. Waqas and M. Imran Cheema\authormark{*} }

\address{Department of Electrical Engineering, Syed Babar Ali School of Science and Engineering, Lahore University of Management Sciences (LUMS), Lahore 54792, Pakistan}
\email{\authormark{*}imran.cheema@lums.edu.pk} 


\begin{abstract*} 
Civil infrastructure is often exposed to high pressures and temperatures, requiring continuous structural health monitoring to ensure safe and reliable operations. Optical fiber sensors are ideal for such environments and can measure multiple physical parameters with a single sensing head. Previous work has demonstrated that optical fiber-based sensors can simultaneously measure multiple parameters. However, these sensing techniques face various challenges, including cross-sensitivity, complex demodulation schemes, fragile structures, and complex fabrication procedures. To address these issues, we present a novel sensing scheme that combines Gaussian process regression~(GPR), a learning-based approach, with phase shift fiber cavity ring down spectroscopy~(PS-CRDS), for the simultaneous measurement of strain and temperature. Our sensor comprises a Fabry-P\'erot cavity formed by two fiber Bragg gratings, with a multimode fiber section inside the cavity serving as the sensing head for both parameters. We apply GPR to the relative phase extracted from PS-CRDS measurements as a single input feature to estimate strain and temperature, using both physics-based and experimental data. For the experimental data, our approach achieves root mean square errors of \SI{0.0094}{\(\celsius\)} for temperature and \SI{34.36}{\(\mu\varepsilon\)} for strain. The experimental temperature and strain sensitivities are \SI{0.049}{\degree/\celsius} and \SI{0.008}{\degree/$\mu\varepsilon$}, respectively. We anticipate that our work will lead to viable multiparameter sensors for various applications not only in structural monitoring but also in agriculture, food, and health sectors.  

\end{abstract*}

\section{Introduction}
Infrastructures such as buildings, bridges, liquid and gas pipelines, railway tracks, and electricity transmission lines are advancing and expanding rapidly. Consequently, there is an increasing need to simultaneously sense multiple physical parameters to assess their structural and environmental health and ensure safe operation. Optical fiber-based sensors are strong candidates for such systems because of their several benefits over conventional sensors. These include immunity to electromagnetic interference, non-conductive nature, resistance to harsh environments, minimal signal propagation loss, and multiplexing capabilities. In addition, optical fiber sensors can also work as versatile sensing systems because they can operate in diverse environments with minimal changes to their sensing structures. 

A single fiber sensing head can provide information on multiple physical parameters, eliminating the need to package several specialized sensors. Consequently, multiparameter sensing using a single optical fiber-based sensing head has attracted considerable interest in recent years~\cite{Wang_2025,9178745,XU2024111293,Liang_2024,Liang_2023,Hu:22}. However, simultaneous monitoring of multiple parameters introduces the cross-sensitivity challenge, i.e., a sensor's response to one measurand is influenced by other measurands~\cite{Huang_2026,Han:23,Li_2026}.

To minimize cross-sensitivity issues in multiparameter sensing, researchers have proposed various modalities in prior work. Some of the notable works include applications of fiber Bragg gratings~(FBG) in various forms~\cite{Huang_2026,Wang_2025,9178745,XU2024111293}, fiber-based interferometers~(FPI)~\cite{Hu:22,Han:23}, and various FPI-FBG cascaded platforms~\cite{Li_2026,Liang_2024,Liang_2023}. Although these approaches report multiparameter sensing with reasonable sensitivity and reduced cross-sensitivity, they generally require additional sensing elements and complex sensing architectures or fabrication protocols. Their performance can also deteriorate when the sensor responses are nonlinear or when the sensitivity matrix becomes poorly conditioned.

Recently, machine learning and deep learning techniques have emerged as alternative approaches for multiparameter sensing. For example, researchers have utilized machine learning~\cite{10606171,9337871,Saha2025}, artificial neural networks~\cite{DEY2022113254}, deep neural networks~\cite{KIKUCHI2022107108}, and a deep learning-based APDNet architecture~\cite{DENG2023103356} with different FBG-based sensing platforms for simultaneous temperature and strain measurements. The data-driven approaches have also been applied to non-FBG-based systems for multiparameter sensing. For example, researchers have employed Gaussian process regression~(GPR) to simultaneously estimate strain and temperature from Brillouin frequency shifts in an SMF-28 fiber sensor~\cite{Karapanagiotis:23}. The work further shows that GPR outperforms various regression methods, including linear least squares, polynomial, ridge, and Bayesian ridge. 

Despite the aforementioned advances, most learning-based approaches for multiparameter fiber sensing rely on wavelength shifts or other features extracted from optical spectra. In contrast, optical cavities provide a multipass sensing platform that can enhance the light interaction with stimuli in the sensing head and, consequently, improve sensitivity. However, the application of machine learning and deep learning to the dynamic response of optical cavities to simultaneously estimate multiple physical parameters remains largely unexplored. 

In this work, we address the aforementioned gaps by combining phase-shift fiber cavity ring-down spectroscopy~(PS-CRDS)~\cite{DaniyalGhauri:21} with Gaussian process regression~(GPR) for simultaneous inverse estimation of strain and temperature. To the best of our knowledge, this is the first demonstration of GPR-assisted simultaneous estimation of strain and temperature using the PS-CRDS response of a fiber cavity.  Our sensing platform exhibits experimental sensitivities for the inverse estimation of strain and temperature from PS-CRDS data of \SI{0.049}{\degree/\celsius} and \SI{0.008}{\degree/$\mu\varepsilon$}, respectively, and our GPR model exhibits the corresponding root mean square errors of \SI{0.0094}{$\celsius$} and \SI{34.36}{$\mu\varepsilon$}, respectively. The cross-sensitivity of our scheme is \SI{0.1632 }{$\mu\varepsilon$/$\celsius$}, which is significantly reduced using a simpler sensing platform compared to previous works. 

We now describe the rest of the paper. In Section ~\ref{sec:theory}, we present the theoretical background of the PS-CRDS technique as a function of strain and temperature, followed by our GPR model description. In Section ~\ref{sec:exp}, we describe the experimental setup for conducting PS-CRDS measurements under applied strain and temperature using a multi-mode fiber sensing head. In Section ~\ref{sec:res}, we present the results produced by the experimental measurements, physics-based modeling, and GPR-based inverse estimation. Finally, we discuss our results and conclude the paper in Section ~\ref{sec:con}.    

\section {Theory}
\label{sec:theory}
Phase-shift cavity ring-down spectroscopy~(PS-CRDS) uses a sinusoidally modulated continuous-wave laser beam as the input to an optical cavity. The cavity output exhibits a phase shift relative to the input. The phase shift is related to the cavity ring-down time, $\tau$, as follows~\cite{ENGELN1996105, Lehmann2009}:
\begin{equation}
\tan(\phi) = -\omega_m\tau
\end{equation}
where $\omega_m$ is the angular modulation frequency. The negative sign indicates that the cavity output lags the input beam. The ring down time, $\tau$, in terms of the effective refractive index~($n$), length~($l$), and the cavity finesse~($\mathcal{F}$) is~\cite{Pedrotti2008,Yariv2007}: 
\begin{equation}
\tau = \dfrac{nl\mathcal{F}}{\pi c}
\end{equation}
where $c$ is the speed of light. The applied strain on the cavity changes its refractive index, length, and finesse. However, the refractive index and length variations due to strain or temperature are negligible~\cite{XU2019163159,8491323} and, hence, have an insignificant effect on the phase shift. Consequently, the strain-dependent phase shift, $\phi(\sigma)$, is given below:
\begin{equation}
\phi (\sigma) = -\tan^{-1}\left(\dfrac{2 f_m nl\mathcal{F}(\sigma) }{c}\right)
\label{eq: t}
\end{equation}
Equation~\eqref{eq: t} can be re-written as:
\begin{equation*}
\phi (\sigma) = -\tan^{-1}\left(\dfrac{2 f_m nl\int(\frac{d\mathcal{F}(\sigma)}{d\sigma})d\sigma }{c}\right)
\end{equation*}
Integrating with respect to strain, $\sigma$, we get the following strain-dependent phase shift:
\begin{equation}
\phi (\sigma) = -\tan^{-1}\left(\dfrac{2 f_m nl(\frac{d\mathcal{F}(\sigma)}{d\sigma}\sigma+\mathcal{F}_\sigma) }{c}\right)
\label{eq:strain}
\end{equation}
Similarly, the temperature-dependent phase shift is given by:
\begin{equation}
\phi (T) = -\tan^{-1}\left(\dfrac{2 f_m nl(\frac{d\mathcal{F}(T)}{dT}T+\mathcal{F}_T) }{c}\right)
\label{eq:temp}
\end{equation}
where $\mathcal{F}_\sigma$ and $\mathcal{F}_T$ are the integration constants corresponding to the cavity finesse under zero strain and at room temperature, respectively. 

\subsection{Multi-Output Gaussian Process Regression Using a Linear Model of Coregionalization}
\label{sec:GPR}

The PS-CRDS relative phase response of an optical fiber sensing cavity depends on the applied strain and temperature. We use multi-output Gaussian process regression~(GPR) to solve the inverse problem of estimating these quantities from the measured PS-CRDS phase response. GPR is a non-parametric probabilistic approach that describes the unknown relationship between inputs and outputs through a distribution over functions~\cite{Rasmussen2006}:

\begin{equation}
P(f|X)=\mathcal{N}(f|\mu,K)
\end{equation}
Here, \(X = [\phi_1, \phi_2, ..., \phi_n ]\) represents the input PS-CRDS phase points, and \(f\) contains the corresponding latent values of the strain and temperature functions, stacked output by output. The mean vector \(\mu=m(X)\) describes their expected values, while the covariance matrix \(K\) describes how these values vary together across PS-CRDS phase points and between the two outputs. For \(n\) PS-CRDS phase points and two outputs, \(f\) has \(2n\) elements and \(K\) has dimensions \(2n\times2n\).

We construct this covariance matrix using a linear model of coregionalization~(LMC), which represents the output functions as weighted combinations of shared and independent latent functions~\cite{alvarez2012}. The resulting matrix-valued covariance function is given by:
\begin{equation}
\mathcal{K}(\phi_i,\phi_j)
=
\sum_{q=1}^{F}B_q k_q(\phi_i,\phi_j)
\end{equation}
Here, \(B_q \in \mathbb{R}^{d \times d}\) is a coregionalization matrix with \(d=2\) for the two outputs. Each \(B_q\) is symmetric and positive semidefinite, while its off-diagonal elements describe the covariance between strain and temperature. The scalar kernel \(k_q(\phi_i,\phi_j)\) describes how this contribution varies between input phase points. For the complete dataset, the covariance matrix is defined as a sum of Kronecker products:
\begin{equation}
K(X,X)=\sum_{q=1}^{F}B_q\otimes K_q,
\qquad
[K_q]_{ij}=k_q(\phi_i,\phi_j)
\end{equation}
In our case, \(F=2\): the first component is a radial basis function~(RBF) kernel, and the second is the sum of a linear and constant kernel. The RBF kernel is
\begin{equation}
k_1(\phi_i,\phi_j)=\sigma_v^210^{\left(
-\dfrac{(\phi_i-\phi_j)^2}{2l^2}
\right)}
\end{equation}
Here, \(\sigma_v^2\) controls the variance of the RBF component, while \(l\) controls its characteristic scale along the phase axis. Both parameters are learned from the training data. This kernel captures smooth variations in the output functions, assigning stronger correlations to nearby PS-CRDS phase points. The second kernel is
\begin{equation}
k_2(\phi_i,\phi_j)=\sigma_l^2\phi_i\phi_j+\sigma_c.
\end{equation}
The linear kernel introduces a linear trend, while the constant kernel allows a phase-independent contribution to the covariance. The parameters \(\sigma_l^2\) and \(\sigma_c\) control the strengths of these contributions and are also learned from the training data.

The outputs at new phase points \(X_*\) can be predicted using GPR given the training data and its mean function. The joint distribution of the observed training outputs \(\textbf{y}\) and the latent test outputs \(\boldsymbol{f_*}\) is given below~\cite{Rasmussen2006}:
\begin{equation}
\begin{bmatrix}
\mathbf{y}\\
\boldsymbol{f_*}
\end{bmatrix}
\sim
\mathcal{N}
\left(
\begin{bmatrix}
m(X)\\
m(X_*)
\end{bmatrix},
\begin{bmatrix}
K+\beta_n^2I & K_*\\
K_*^{\top} & K_{**}
\end{bmatrix}
\right)
\qquad
\end{equation}
Here, \(K=K(X,X)\), \(K_*=K(X,X_*)\), \(K_{**}=K(X_*,X_*)\), and $\beta_n^2$ is a noise level for making $K$ positive semi-definite. Conditioning this joint Gaussian distribution on the observed training outputs gives the predictive distribution~\cite{Rasmussen2006}:
\begin{equation}
(\boldsymbol{f_*}|\mathbf{y},X,X_*)
\sim
\mathcal{N}
\left(
K_*^{\top}K^{-1}\mathbf{y},
K_{**}-K_*^{\top}K^{-1}K_*
\right)
\end{equation}
The predictive mean provides the estimated strain and temperature, while the predictive covariance quantifies their uncertainty and correlation under the fitted model. The following algorithm evaluates the predictions and log marginal likelihood using Cholesky factorization~\cite{Rasmussen2006}:
\begin{enumerate}
\setlength{\leftmargin}{0pt}
\setlength{\itemindent}{-4pt}
\setlength{\labelsep}{0.3em}
\setlength{\itemsep}{0pt}
\setlength{\parskip}{0pt}
\renewcommand{\labelenumi}{\arabic{enumi}.}

\item $L = \text{cholesky}(K+\beta_n^2I)$
\item $ \boldsymbol{\alpha}= L^{-\top}(L^{-1}\mathbf{y})$
\item $\boldsymbol{\bar{f}_*} = K_*^{\top}\boldsymbol{\alpha} \qquad \text{predictive mean}$
\item $\mathbf{v} = L^{-1}K_*$
\item $V[\boldsymbol{f_*}] = K(X_*,X_*)-\mathbf{v}^{\top}\mathbf{v} \qquad \text{predictive variance}$
\item $
\log p(\mathbf{y}|X)
=
-\frac{1}{2}\mathbf{y}^{\top}\boldsymbol{\alpha}
-\frac{1}{2}\log\det(K+\beta_n^2I)
-\frac{n}{2}\log 2\pi
$
\item $\textbf{return}: \boldsymbol{\bar{f}_*} \; (\text{mean}),
V[\boldsymbol{f_*}] \; (\text{variance}),
\log p(\mathbf{y}|X)
\; (\text{log marginal likelihood})$
\end{enumerate}

The log marginal likelihood can be maximized to learn the kernel and coregionalization parameters. Once the mean and variance functions are determined, one can also infer values outside the dataset. Also, it shows that the predictive mean is a linear combination of the outputs and that the predictive variance depends only on the inputs.

\section{Experimental Setup}
\label{sec:exp}
In this section, we describe the temperature and strain setups that are used to apply both stimuli to the sensing head simultaneously. We also describe the overall PS-CRDS experimental setup. 

\subsection{Temperature and Strain Setups}
The temperature and strain setups provide controlled temperature and longitudinal strain, respectively, to a section of multimode fiber (MMF) that serves as the sensing head. The temperature setup consists of a flexible resistive foil heating element (HT10 K) of size 25.4~mm $\times$ 76.2~mm with an integrated 10~k$\Omega$ NTC thermistor for temperature sensing, together with a TC200 controller, all from THORLABS Inc. The fiber is placed at the center of the heater and aligned along its length. The controller monitors the fiber surface temperature using the thermistor at the center of the heater and regulates it by adjusting the heater input voltage via a proportional-integral-derivative (PID) control loop. However, due to silica's low thermal conductivity~(1~W/m.K), changes in the  MMF surface temperature produce very small changes in the PS-CRDS output signal.  To enhance the thermal response, we coat the MMF with a 20~nm-thick gold layer using physical vapor deposition. A thin chromium pre-layer (5~nm) is also deposited to improve adhesion between the gold layer and the fiber.

The strain setup consists of two stages: a fixed stage and a movable stage. The fixed stage is constructed using two posts and a plate. We position the fiber between the two stages and secure it using an ultraviolet-curable glue. The movable stage is then used to apply longitudinal strain to the fiber. 

\subsection{Experimental Setup of PS-CRDS}
\begin{figure}[H]
\centering
\includegraphics[width=13cm]{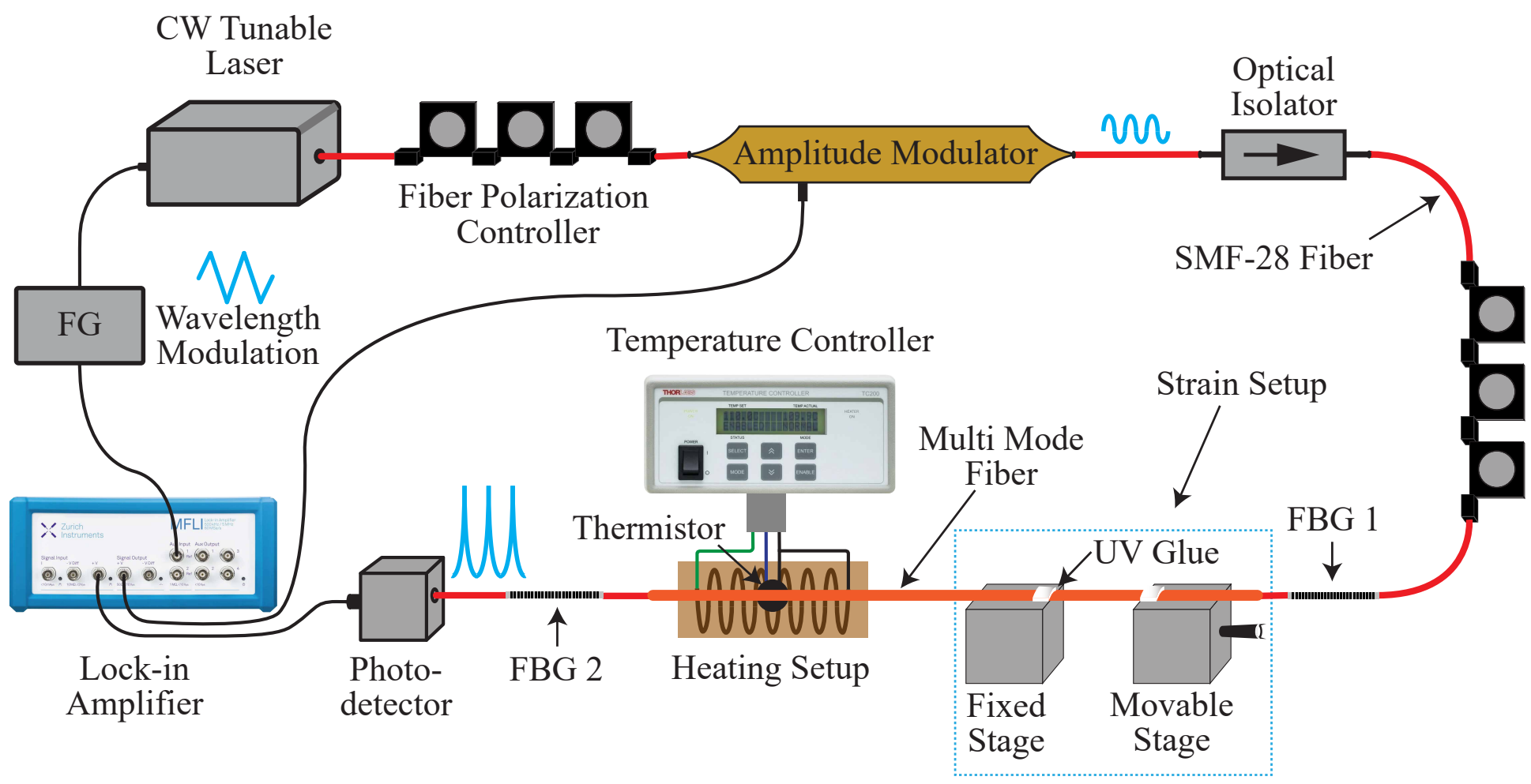}
\centering
\caption{Schematics of a sensor for measuring strain and temperature simultaneously on the single sensing head using PS-CRDS.}
\label{fig:experimental_setup}
\end{figure}

The schematic diagram of the overall experimental setup is shown in Fig.~\ref{fig:experimental_setup}. We build an optical fiber-based cavity by splicing fiber Bragg gratings~(FBGs) at both ends of the MMF sensing head. 
The length of the MMF section is 1.94~m, and the overall cavity length is 5.85~m. FBG~1 and FBG~2 have a central wavelength of 1550~nm and a reflectivity of 95\%. Their 3-dB bandwidths are 0.38~nm and 0.28~nm, respectively. We use a 1550~nm continuous-wave single-frequency laser diode, operated with a CLD1015 controller (both from THORLABS Inc.), as the sensor source. To scan the cavity resonance peaks via wavelength modulation, the laser is current-modulated using a 10~Hz, 50~mV$_{pp}$ ramp signal applied to the controller's RF port by a function generator. 

For the PS-CRDS measurements, we sinusoidally modulate the laser output at 3~MHz using an external Mach-Zehnder modulator~(Agiltron LNML-122151113). A fiber polarization controller~(Thorlabs FPC562) is used to ensure that the modulator operates at the optimal polarization.  The optical isolator prevents the modulator from receiving the back-reflected light from the cavity. A second fiber polarization controller~(Thorlabs FPC032) is used to select either the TE or TM cavity mode. The wavelength- and intensity-modulated laser light is injected into the cavity, and the cavity output is monitored by an InGaAs photodetector~(Thorlabs DET01CFC/M). The detector output is fed to the lock-in amplifier~(Zurich Instruments MFLI), which measures the phase shift of the cavity output relative to the modulated input.

\section{Results}  
\label{sec:res}
\begin{figure}[H]
\centering
\includegraphics[width=8cm]{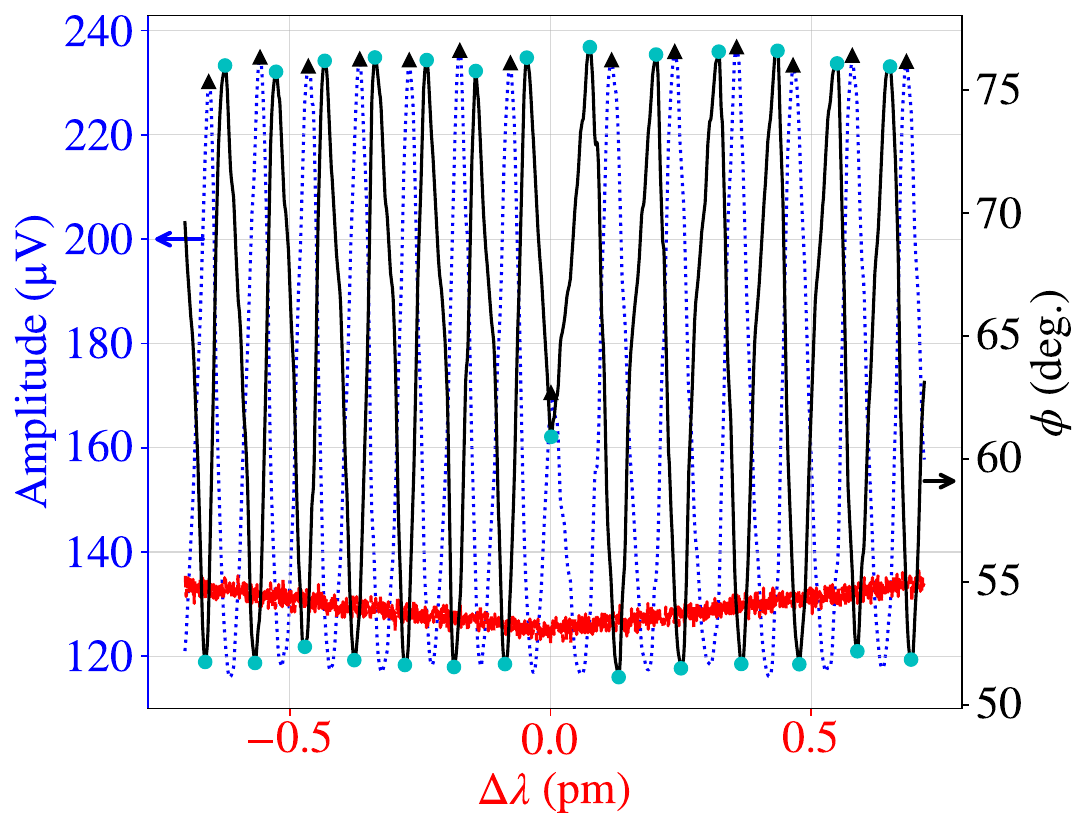}
\caption{FP-cavity resonance peaks~(blue), PS-CRDS phase dips~(black), and the wavelength modulation signal~(red).}
\label{fig:phase_amp_signal}
\end{figure}

\begin{figure}[H]
\centering\includegraphics[width=12.5cm]{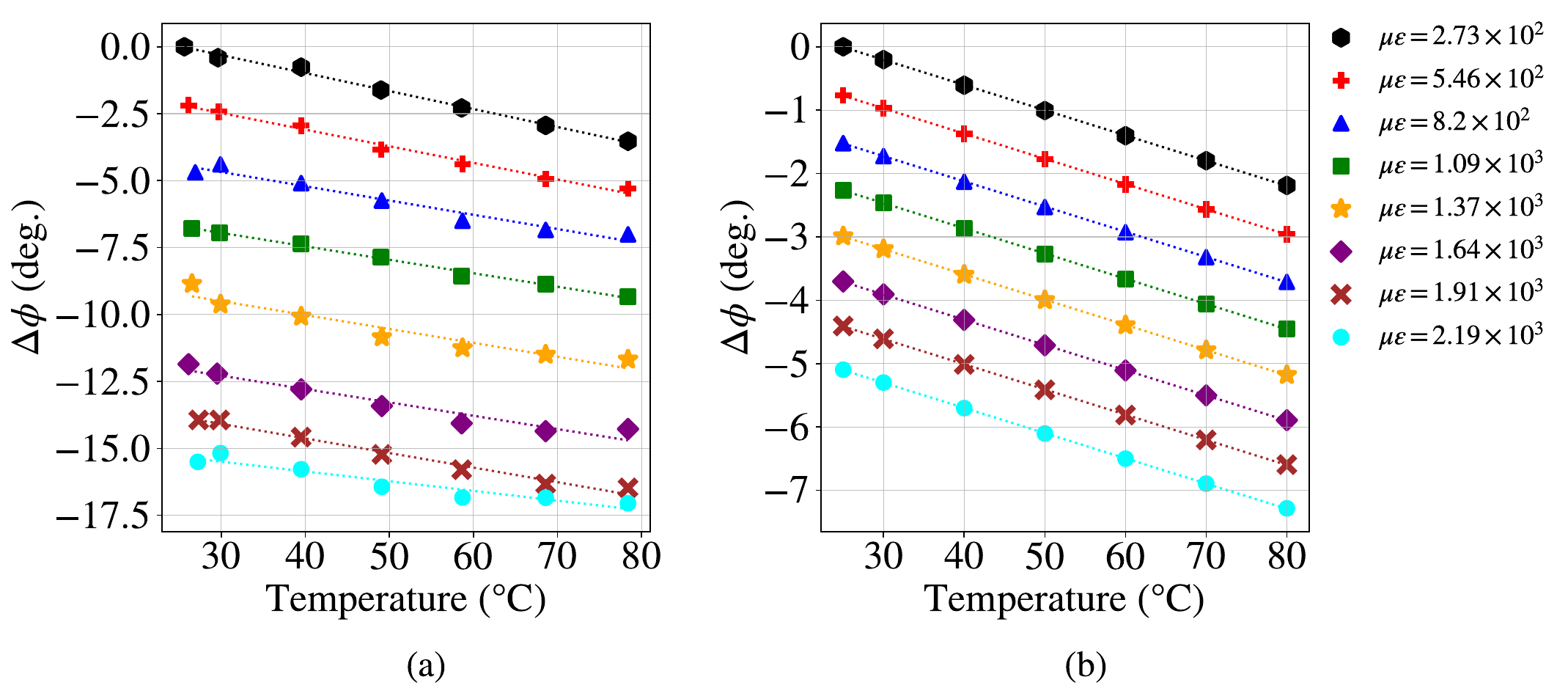}
\centering\caption{Relative PS-CRDS phase difference versus stimuli. (a) Experimental relative phase difference versus temperature; each fitted line represents a constant strain given in the sidebar (b) Theoretical results}
\label{fig:data}
\end{figure}

The PS-CRDS measurements yield a phase shift spectrum containing dips at the cavity resonances, as shown in Fig.~\ref{fig:phase_amp_signal}. The resonance dips provide the phase shift of the cavity output relative to the sinusoidal input. The theoretical origins of the resonance phase shift spectrum are described in our earlier work~\cite{DaniyalGhauri:21}. 
Applying temperature and strain to the sensing head changes the measured phase shift, i.e., the depth of the dips. To determine the phase shift associated with a given change in temperature or strain, we use the following method. For each applied temperature or strain value, the phase spectrum is recorded 10--12 times. Each spectrum contains approximately 70--75 resonance dips, and the average phase shift of these resonance dips is taken as the phase shift value for that spectrum. The phase shift values obtained from the 10--12 spectra are then averaged to determine a single absolute phase shift value for the corresponding temperature or strain condition. Finally, the relative phase shift, $\Delta\phi$, is calculated by subtracting the phase shift value at the initial stimulus condition from all subsequent values, such that the response begins at zero degrees, as shown in Fig.~\ref{fig:data}.

In Fig.~\ref{fig:data}a, we acquire the data from PS-CRDS measurements as follows. We fix the strain and vary the temperature from approximately \SI{25}{\celsius} to \SI{80}{\celsius}. At each temperature, we wait approximately 3--5 minutes for the temperature to reach its desired value, stabilize, and produce a corresponding change in the phase shift dip. We repeat the same procedure for each strain value, ranging from  273 $\mu\varepsilon$ up to 2190 $\mu\varepsilon$. The corresponding theoretically predicted relative phase shift values, shown in Fig.~\ref{fig:data}b, are calculated for the experimentally applied temperature and strain values using Eqs.~\eqref{eq:strain} and \eqref{eq:temp}. Both datasets exhibit similar trends, demonstrating good agreement between the experimental observations and the underlying theoretical model. 
The experimentally measured sensitivities of temperature and strain are \SI{0.049}{\degree/\celsius} and \SI{0.008}{\degree/$\mu\varepsilon$}, respectively, whereas the corresponding theoretical sensitivities are \SI{0.04}{\degree/\celsius} and \SI{0.003}{\degree/$\mu\varepsilon$}, respectively.
We can clearly see that Eqs.~\eqref{eq:strain} and \eqref{eq:temp} require the values of $\mathcal{F}_\sigma$ (cavity finesse at zero strain), $\mathcal{F}_T$ (cavity finesse at room temperature), $\frac{d\mathcal{F}(\sigma)}{d\sigma}$ (finesse variation with strain), and $\frac{d\mathcal{F}(T)}{dT}$ (finesse variation with temperature) to generate the theoretical plots shown in Fig.~\ref{fig:data}b.  

\begin{figure}[H]
\centering
\includegraphics[width=12.5cm]{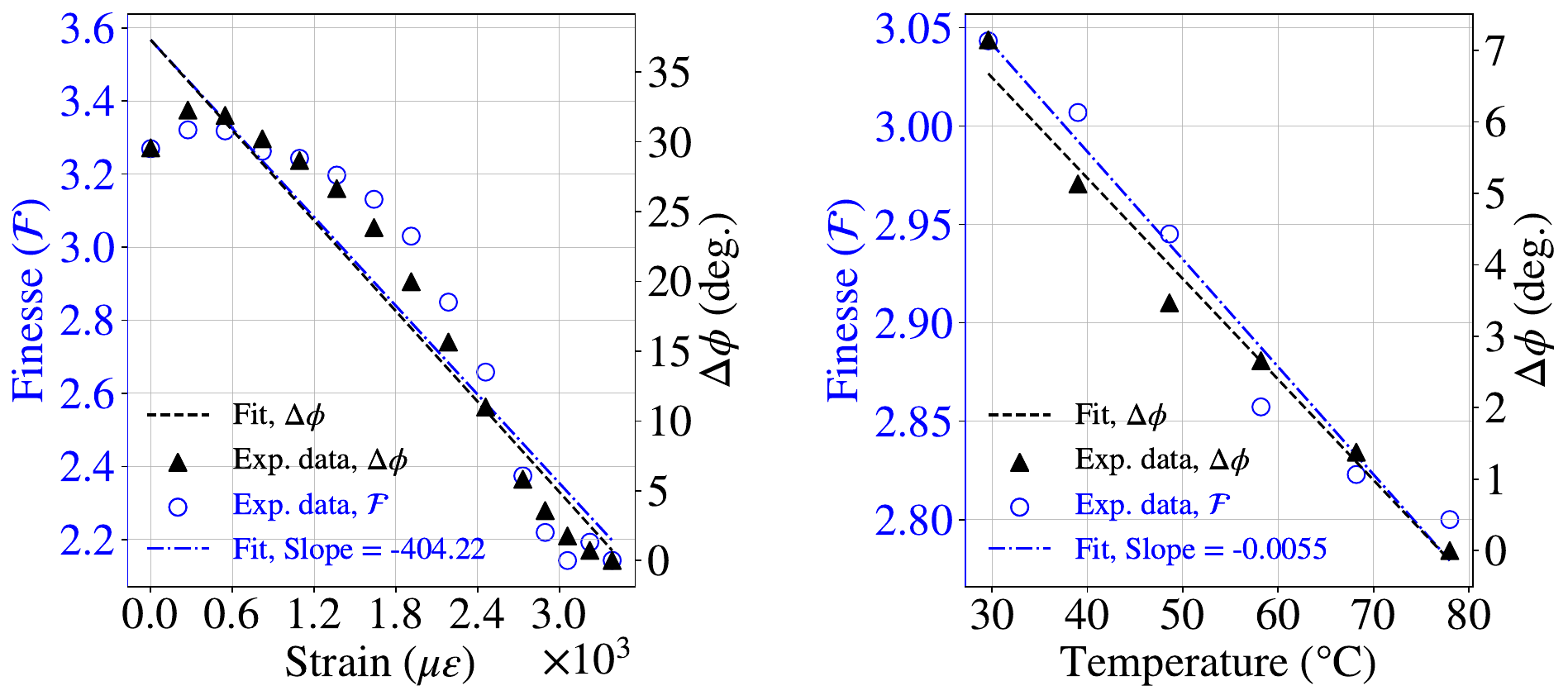}
\centering
\caption{Stimuli effect on cavity finesse and relative output PS-CRDS phase difference. (a) Strain effect (b) Temperature effect. }
\label{fig:finesse_info}
\end{figure}

\begin{figure}[H]
\centering
\includegraphics[width=10cm]{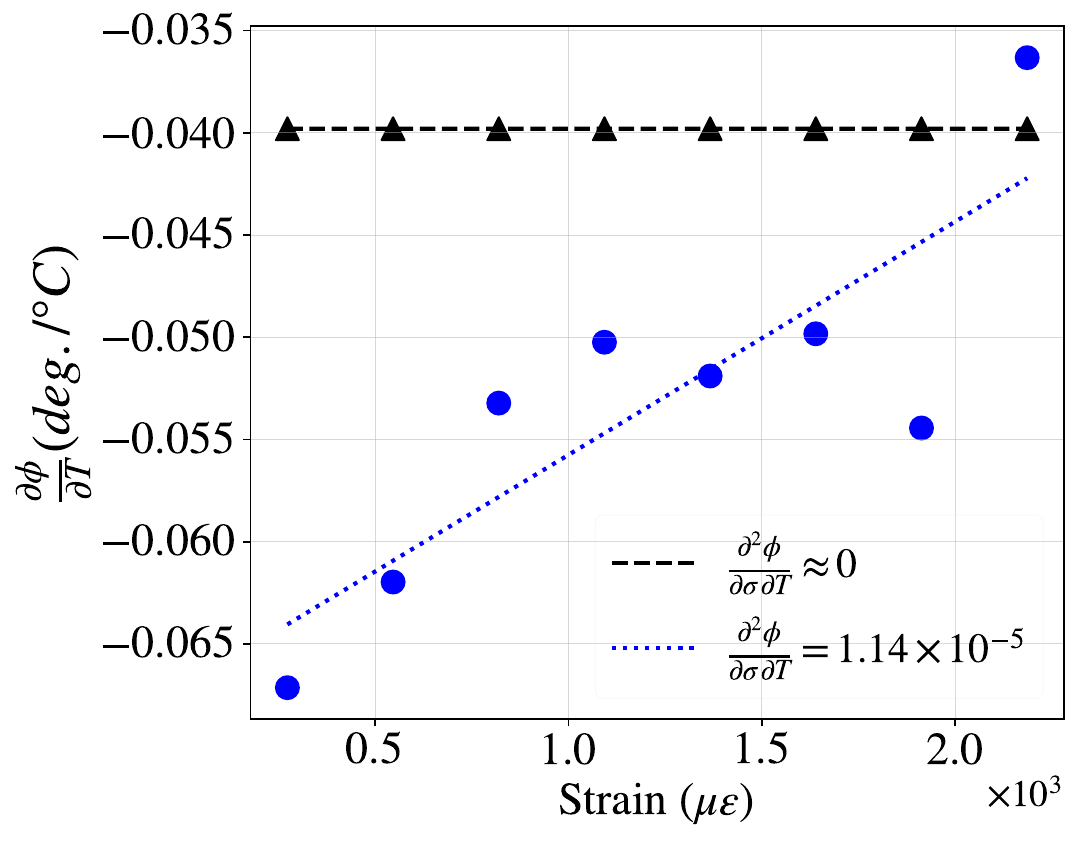}
\centering
\caption{Temperature sensitivity, \(\frac{\partial\phi}{\partial T}\), as a function of applied strain, \(\sigma\). Black triangles and blue circles represent the theoretical model predictions and experimental results, respectively. The dashed and dotted lines show their corresponding linear fits. The fitted slopes quantify the temperature–strain interaction, \(\frac{\partial^2\phi}{\partial\sigma\,\partial T}\).}
\centering
\label{fig:temp_strain_intraction_coefficient}
\end{figure}

Therefore, we conduct separate PS-CRDS measurements for strain and temperature to determine the corresponding phase shift and finesse variations, as shown in Fig. \ref{fig:finesse_info}(a) and (b), respectively. The quantities, $\mathcal{F}_\sigma$ and $\mathcal{F}_T$, are y-intercepts of the fitted lines, whereas  $\frac{d\mathcal{F}(\sigma)}{d\sigma}$ and $\frac{d\mathcal{F}(T)}{dT}$ correspond to their slopes.
We also investigate how our sensor's temperature sensitivity is affected by applying different strains to the sensing head. It is evident from Fig.~\ref{fig:temp_strain_intraction_coefficient} that the theoretical temperature sensitivity is almost independent of the strain. However, the experimental temperature sensitivity depends very weakly on the applied strain. 

To solve the inverse problem of determining temperature and strain from PS-CRDS phase-shift measurements, we employ the GPR model described in Section~\ref{sec:GPR}. We split the data into training~(80\%) and test~(20\%) sets, with the former used to train the model and the latter used to evaluate its performance on unseen data. Before feeding the data to the model, we normalize it and rotate it using a two-dimensional transformation matrix to minimize unnecessary correlations in the coregionalization matrix of the multi-output GPR model. After training, the original variables are recovered using the corresponding inverse transformation.

\begin{figure}[H]
\centering
\includegraphics[width=12.5cm]{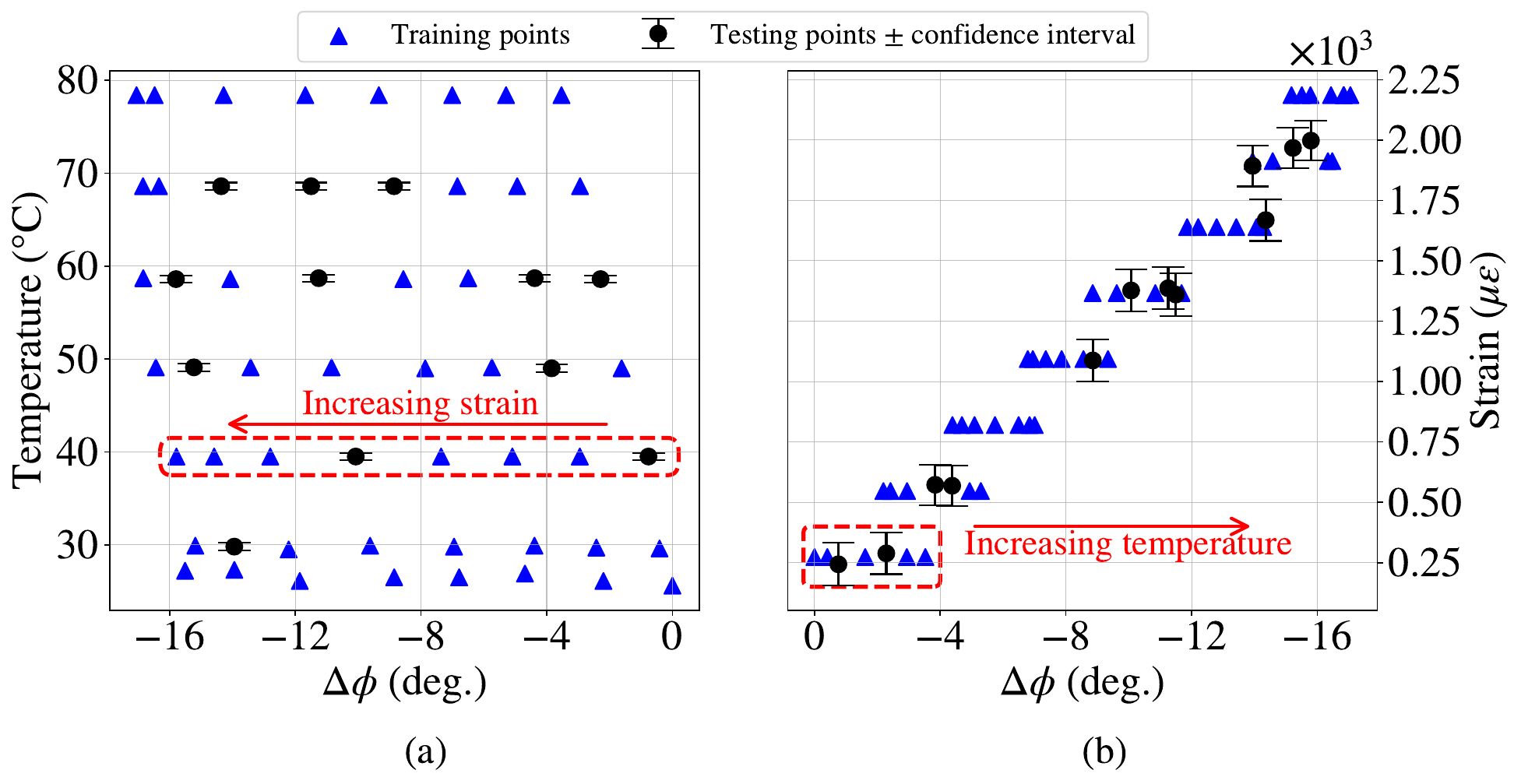}
\caption{GPR model output for experimental data. (a) Temperature versus PS-CRDS phase with varying strains. Each rightmost and leftmost point in every row is at $0.25 \times 10^3$ and  $ 2.25\times 10^3$, respectively, as represented by an arrow on one row. (b) Strain versus PS-CRDS phase with varying temperatures. Each leftmost and rightmost point in every row is approximately at 25$^o$C and  80$^o$C, respectively, as represented by an arrow on one row.}
\label{fig: exp_result}
\end{figure}

\begin{figure}[H]
\centering
\includegraphics[width=12.5cm]{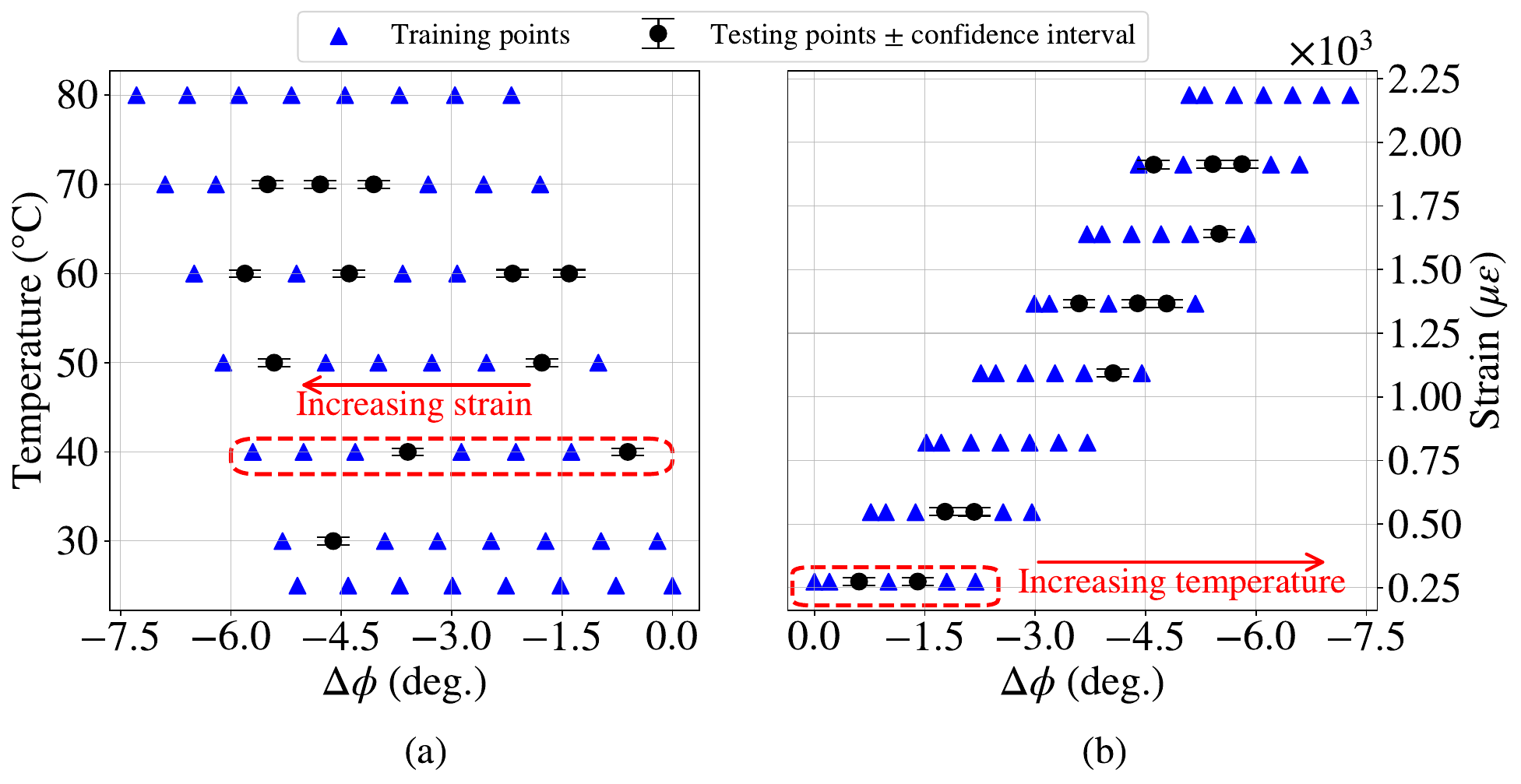}
\caption{GPR model output for physics-based data. (a) Temperature versus PS-CRDS phase with varying strains. (b) Strain versus PS-CRDS phase with varying temperatures.}
\label{fig: phy_result}
\end{figure}

We use the negative log marginal likelihood~(NLL) as the loss function and a multitask kernel comprising a radial basis function~(RBF) kernel with a base-10 and an additive kernel of linear and constant functions.  We train the model for 5000 epochs with a learning rate of 0.025 on each dataset separately. The optimized NLL values are -3.344 and -2.490 for the physics-based and experimental datasets, respectively. The corresponding results are shown in Fig.~\ref{fig: exp_result} and Fig.~\ref{fig: phy_result} for the experimental and physics-based datasets, respectively.
For the physics-based data, the root-mean-square errors~(RMSEs) of the predicted test points are \SI{0.0002}{$\celsius$} for temperature and \SI{1.397}{$\mu\varepsilon$} for strain. For the experimental data, the corresponding RMSEs are \SI{0.0094}{$\celsius$} and \SI{34.36}{$\mu\varepsilon$}, respectively.

\section{Conclusion}
\label{sec:con}
In this work, we demonstrate the first application of GPR to the inverse estimation of strain and temperature from PS-CRDS measurements in a fiber cavity sensor. Our experimental results show that the output PS-CRDS phase varies linearly with changes in temperature and strain. The physics-based predictions of the phase response to these stimuli agree well with the experimental results, as shown in Fig. \ref{fig:data}.  

Unlike previously demonstrated multiparameter optical fiber sensors, which may suffer from cross-sensitivity, complex calibration, intensity degradation, birefringence, and the need for customized sensing structures, our sensing modality offers a simpler architecture and interrogation scheme. Moreover, it can be readily extended to additional sensing parameters without altering the overall setup or learning framework. Table~\ref{tab_comparison} compares our sensor with previous work on multiparameter sensing using machine learning or deep learning. 

\begin{table}[H]
\centering
\caption{Comparison of learning-based approaches for simultaneous
temperature and strain sensing.}
\label{tab_comparison}
\small
\renewcommand{\arraystretch}{1.15}
\setlength{\tabcolsep}{3pt}

\begin{tabular}{|
>{\centering\arraybackslash}p{2.0cm}|
>{\centering\arraybackslash}p{2.0cm}|
>{\centering\arraybackslash}p{2.5cm}|
>{\centering\arraybackslash}p{1.9cm}|
>{\centering\arraybackslash}p{3.7cm}|}
\hline

\textbf{Work} &
\textbf{Platform} &
\textbf{Interrogation instrument} &
\textbf{Learning method} &
\textbf{Performance} \\
\hline

Sarkar \textit{et al.} (2021)~\cite{9337871} &
Single FBG &
Commercial FBG interrogator (SM125) &
Ensemble ML &
$\sim$90\% prediction accuracy \\
\hline

Dey \textit{et al.} (2022)~\cite{DEY2022113254} &
Etched FBG + SMS &
Optical spectrum analyzer (OSA) &
ANN &
RMSE: 0.025 $^\circ$C, 0.195 $\mu\varepsilon$ \\
\hline

Kikuchi \textit{et al.} (2022)~\cite{KIKUCHI2022107108} &
Tilted FBG &
Tunable-laser spectral interrogation &
Deep NN / SVR &
Best RMSE: 1.49 $^\circ$C, 25.9 $\mu\varepsilon$ \\
\hline

Deng \textit{et al.} (2023)~\cite{DENG2023103356} &
FBG sensing system &
Commercial FBG interrogator &
ADPNet &
RMSE: 0.73 $^\circ$C, 141.76 $\mu\varepsilon$ \\
\hline

Karapanagiotis \textit{et al.} (2023)~\cite{Karapanagiotis:23} &
BOFDA, SMF-28 &
Vector network analyzer (VNA) &
GPR &
MAE: 2 $^\circ$C, 45 $\mu\varepsilon$ \\
\hline

Choi \textit{et al.} (2024)~\cite{10606171} &
Single FBG &
Wavelength-swept laser system &
Machine learning &
$R^2 \approx 0.952$ \\
\hline

\textbf{Current work} &
\textbf{PS-CRDS fiber cavity} &
\textbf{No interrogator/OSA/VNA} &
\textbf{Multi-output GPR} &
\textbf{RMSE: 0.0094 $^\circ$C, 34.36 $\mu\varepsilon$} \\
\hline

\end{tabular}
\end{table}

We find that our experimental results exhibit higher RMSEs than the physics model, suggesting that our model may not fully capture experimental nonidealities. Possible future directions include further improving the agreement between the model and the experiments, extending the approach to broader operating ranges and dynamically varying stimuli, and evaluating its robustness to additional perturbations. Significantly, our sensing modality can also be extended to different sensing heads and a larger number of other cross-sensitive measurands, offering a potential route toward compact multiparameter fiber sensing systems for a wide range of structural, environmental, and biosensing applications.
\FloatBarrier
\section*{Acknowledgments}
The authors are sincerely grateful to Dr. Muhammad Tahir (Associate Professor, Department of Electrical Engineering, LUMS) for his valuable suggestions and feedback, which helped improve this work.

\section*{Funding}
Higher Education Commission of Pakistan~(HEC-NRPU: No. 20-15820)

\section*{Disclosures}
The authors declare no conflicts of interest.

\section*{Data availability}
Data underlying the results presented in this paper are not publicly available at this time but may be obtained from the authors upon reasonable request.

\bibliography{references}

\end{document}